\documentclass[3p,times]{elsarticle}

\usepackage{ecrc}

\volume{00}

\firstpage{1}

\journalname{Nuclear Physics A}

\runauth{J.Zhao}

\jid{nupha}

\usepackage{amssymb}

\usepackage{lineno}

\usepackage[figuresright]{rotating}

\begin{document}

\begin{frontmatter}




\title{Dielectron production from $\sqrt{s_{NN}}$ = 200 GeV Au + Au collisions at STAR}

\author[SINAP,LBL]{Jie Zhao (for the STAR collaboration)}

\address[SINAP]{Shanghai Institute of Applied Physics, Chinese Academy of Sciences, Shanghai 201800, China}
\address[LBL]{Lawrence Berkeley National Laboratory, Berkeley, CA 94720, USA}
\ead{zhaojie@sinap.ac.cn}

\begin{abstract}
We present the first STAR dielectron measurement in 200 GeV Au + Au collisions. Results are compared to hadron decay cocktail to search for in-medium modification of vector mesons in low mass region and thermal radiation in the intermediate mass region. The transverse mass slope parameters in the intermediate mass region are also discussed.  
\end{abstract}


\end{frontmatter}


\section{Introduction:}
Dilepton production has been proposed to serve as a penetrating probe for the hot and dense medium created in high-energy nuclear collisions. Their small final-state interaction cross sections let dileptons escape the interaction region undistorted. Since  dileptons originate from all stages of a heavy ion reaction, their sources vary with the kinematic phase space under consideration: In the low mass region (LMR: mass$<$1.1GeV/$c^{2}$), vector mesons and direct photons are the dominating source, while dileptons in the intermediate mass region (IMR: 1.1$<$mass$<$3GeV/$c^{2}$) primarily stem from Quack-Gluon Plasma(QGP) thermal radiation and semileptonic decays of charmed mesons. In the high mass region (HMR: mass$>$3 GeV/$c^{2}$), heavy quark decays and Drell-Yan processes contribute the most to the dilepton spectrum.  Due to the time-energy correlation, the higher the dilepton pair mass, the earlier the production. Therefore the dilepton distributions, especially in the IMR and HMR, provide  information on early collision dynamics in heavy ion collisions.

With the installation of the full barrel Time-Of-Flight (TOF) detector\cite{TOF}, the electron identification has been significantly improved at STAR, especially in the low momentum region. In this paper, we will present the first STAR results on dielectron production in Au + Au collisions at $\sqrt{s_{NN}}$ = 200 GeV.

\section{Analysis}
The results are obtained from 200 GeV Au+Au collisions, taken in 2010. The main subsystems of the STAR detector\cite{TPC} used for this analysis are the Time Projection Chamber (TPC) and the TOF with 2${\pi}$  azimuthal coverage at mid-rapidity.

\begin{figure}[htbp!]
\centering 
\includegraphics[width= 7.0cm]{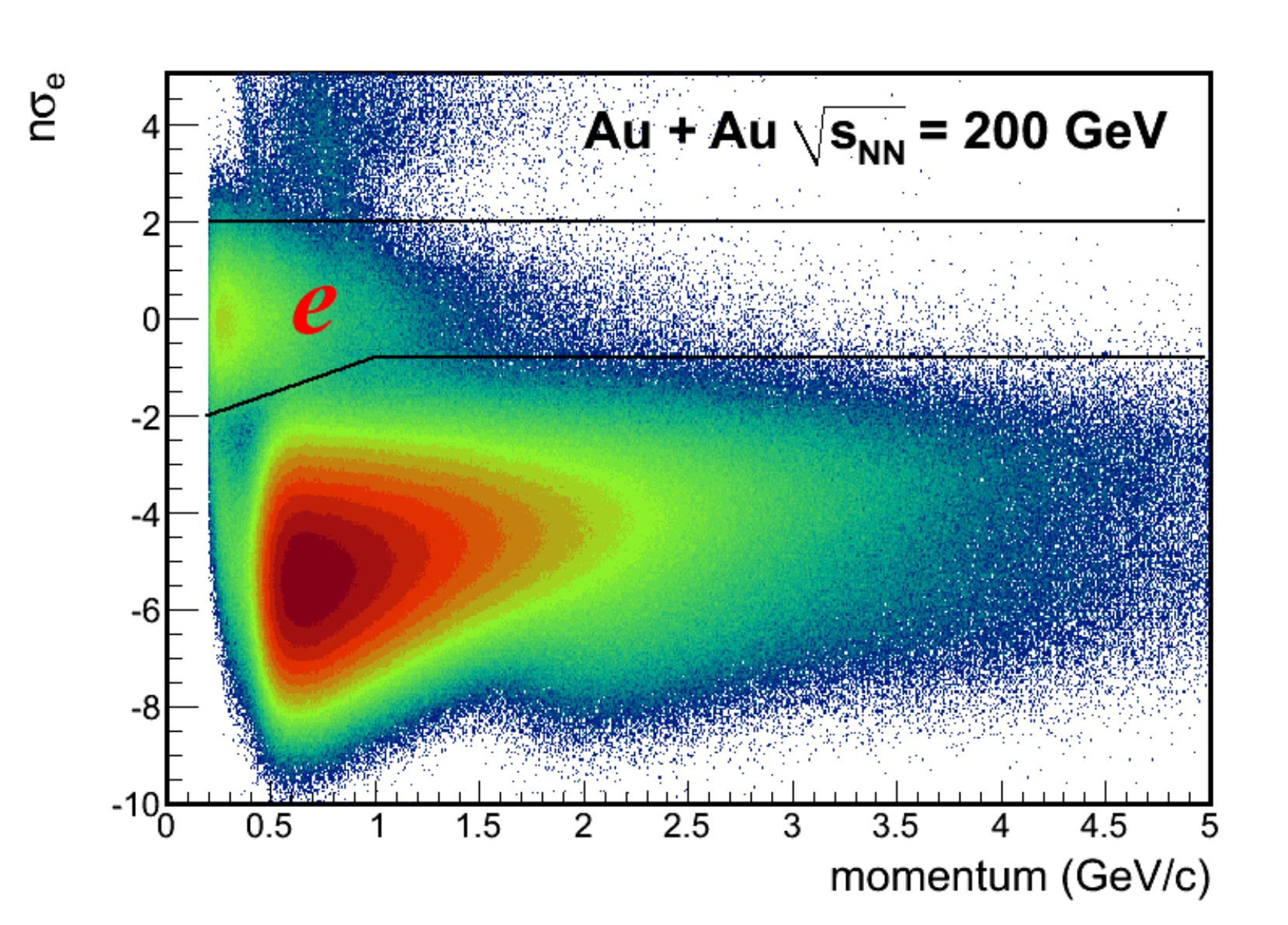} 
\caption{ TPC normalized $dE/dx$ information n$\sigma_{e}$ \cite{TPC} vs momentum with the TOF velocity($\beta$) cut. The two black lines are used to select electrons.} 
\label{EID} 
\end{figure}

In addition to track detection and momentum determination capabilities, the TPC provides particle identification for charged particles by measuring their ionization energy loss ($dE/dx$) in the TPC gas\cite{TPC}. The newly installed full barrel TOF detector provides the particle velocity ($\beta$) information\cite{TOF}. In this analysis, we combine TPC $dE/dx$ and TOF $\beta$ information to identify electrons (Fig.\ref{EID}). The electron purity is $\sim 97\%$ in minimum-bias Au + Au analysis.

\begin{figure}[htbp!]
\centering 
\includegraphics[width= 6.8cm]{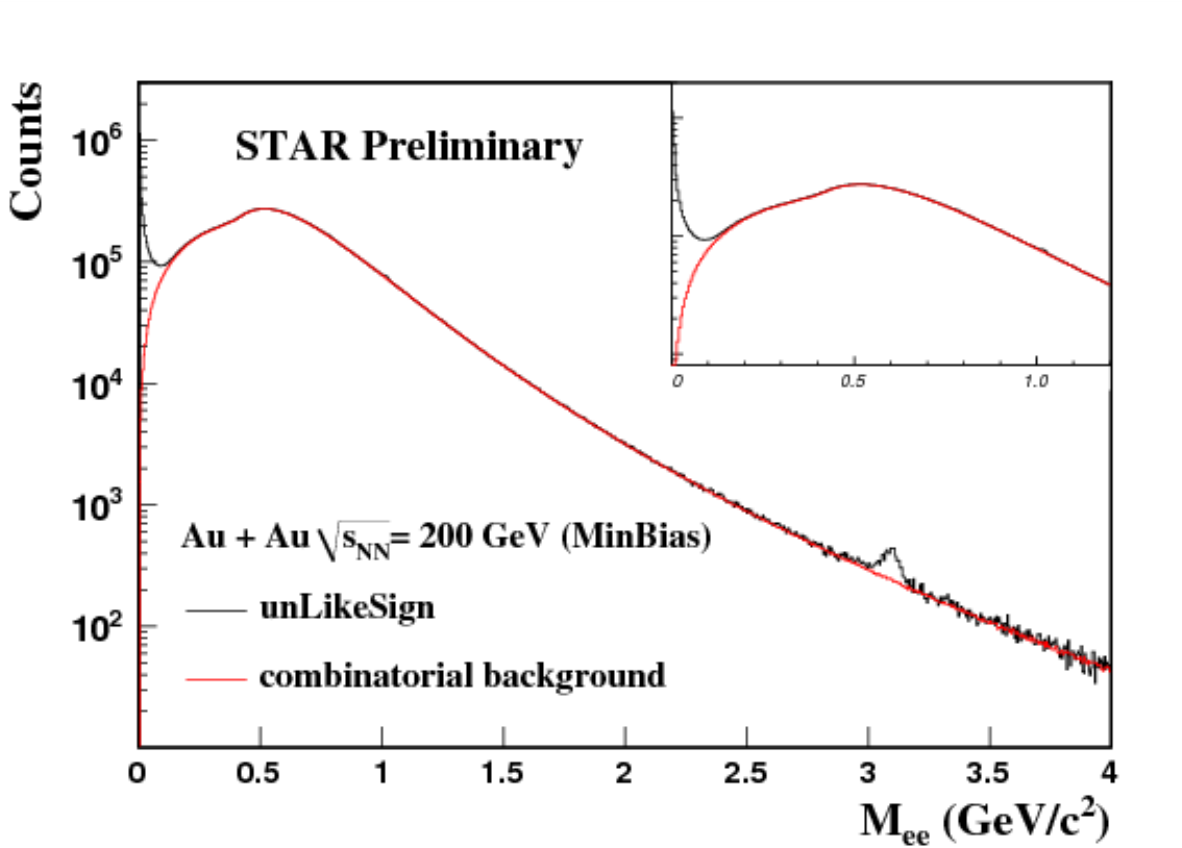} 
\includegraphics[width= 6.8cm]{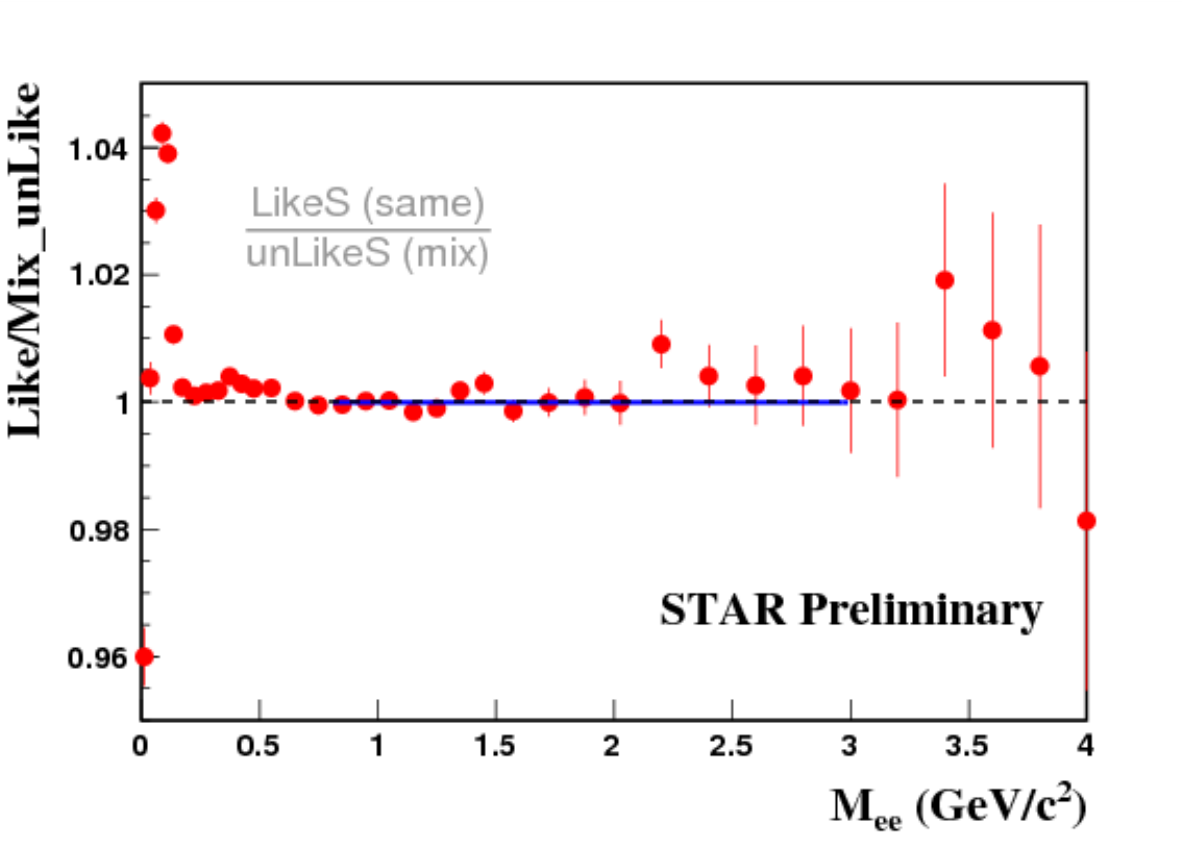}
\caption{Unlike-sign distribution in same- and mixed- event (left), and like-sign (same-event) to unlike-sign (mixed-event) ratio (right).} 
\label{background} 
\end{figure}

The background is subtracted with the mixed-event and like-sign methods. The invariant mass distribution from event mixing agrees with the same-event unlike-sign distribution in the IMR and HMR as shown in Fig.\ref{background} (left). The right plot shows the ratio of the like-sign and mixed-event background. In the LMR, due to the correlated background, from {\it e.g.} cross pair and jet contribution\cite{PHEdie}, we subtract the like-sign background with an acceptance correction. The acceptance correction is needed because the acceptance of like sign pairs is slightly different from that of unlike sign pairs. In the IMR, we use the mixed-event background\cite{PHEdie} since it offers better statistics. The systematic error in the LMR is dominated by the acceptance uncertainty ($< 0.1\%$). In the IMR and HMR, it is dominated by the normalization factor uncertainty from different normalization methods and the normalization region ($< 0.1\%$). 

The dielecton continuum results in this paper are within STAR acceptance ($p_{T}^{e} >$ 0.2 GeV/$c$ , $|\eta^{e}| < 1.0$ , $|y^{ee}| < 1.0$ ) and corrected for efficiency.   

\section{Results}
\begin{figure}[htbp!]
\centering 
\includegraphics[width=5.9cm]{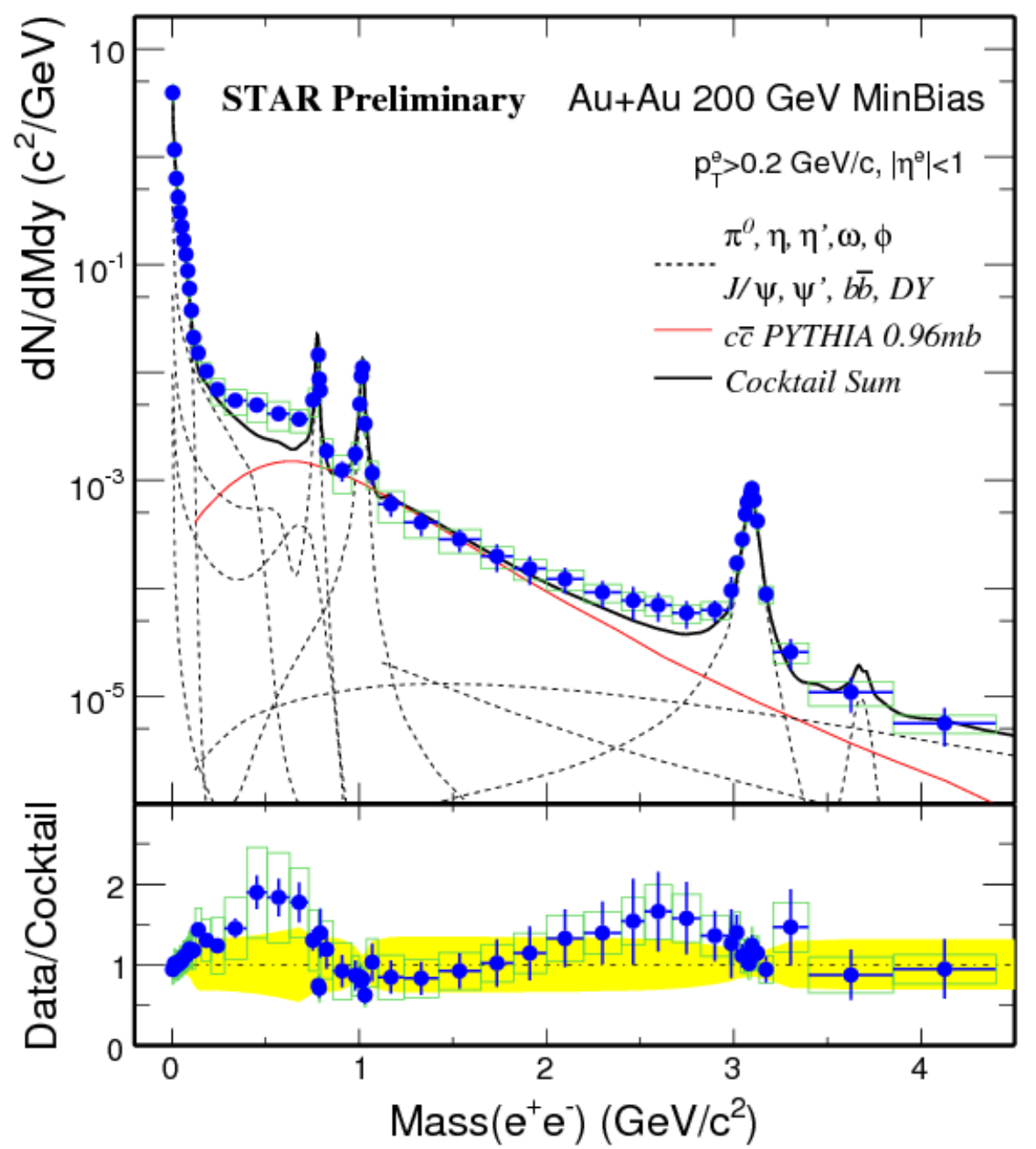} 
\includegraphics[width= 5.9cm]{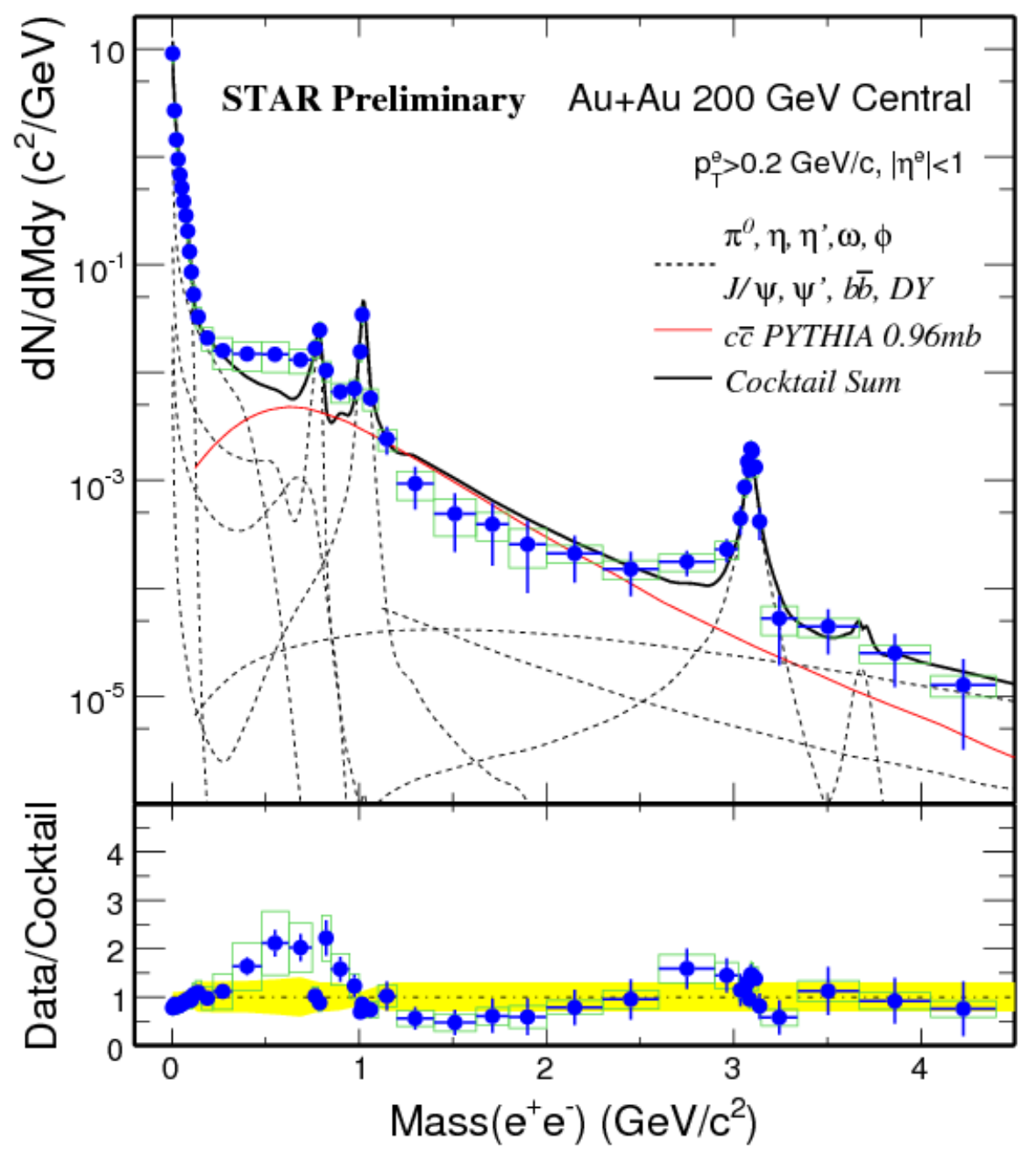} 
\caption{Invariant mass spectra from sqrt{sNN} = 200 GeV Au + Au minimum-bias (left) and central collisions (right). The yellow band is the systematic error on cocktail, while the green box is the systematic error on data.}
\label{AuAu} 
\end{figure}

\begin{figure}[htbp!]
\centering 
\includegraphics[width= 5.0cm]{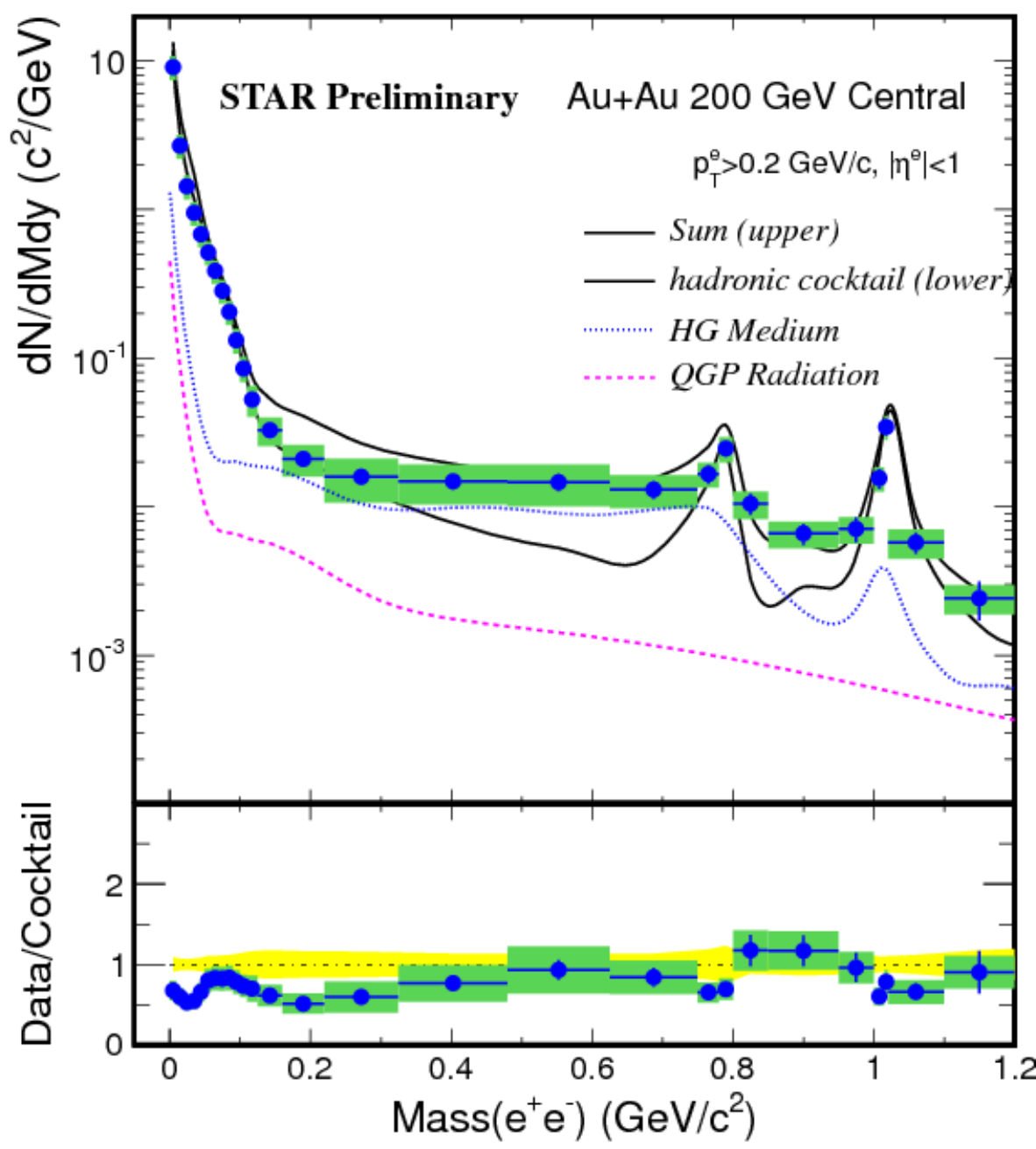} 
\includegraphics[width=7.5cm]{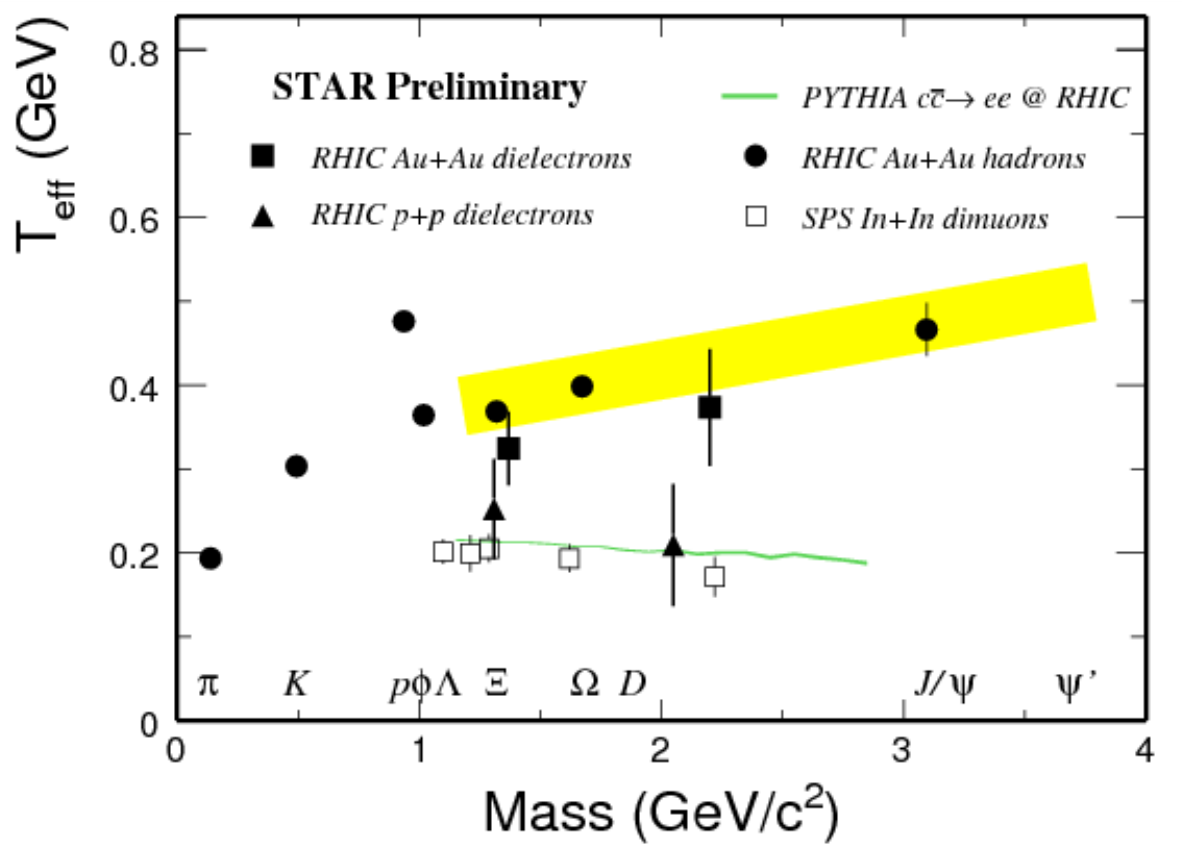} 
\caption{(left) Comparison to the theoretical calculation \cite{Rapp}: the blue and pink dotted lines are the the hadronic gas with medium modifications and QGP radiation separately, the lower black line is the hadronic cocktail, the upper one is sum of the hadronic cocktail with hadronic gas medium modification and QGP radiation contribution, $\sigma_{c\bar{c}} = 0.62 mb$ used in this plot. (right) Transverse mass slope parameters from RHIC and SPS measurements.}
\label{slope} 
\end{figure}

Fig.\ref{AuAu} shows the invariant mass spectra from 200 GeV Au + Au minimum-bias (left) and central collisions (right). The slope parameters at intermediate mass can be found in Fig.\ref{slope}\cite{Huang,NA60,STARh, PHEh}. 

The results from p + p collisions are consistent with hadron decay cocktail simulations\cite{QM2011, ppPaper}, which provide a baseline for Au + Au collisions.

In the LMR we observe an enhancement with respect to the cocktail of $1.72 \pm 0.1    0^{stat} \pm 0.50^{sys} $ (without $\rho$) for central Au-Au collisions, while the enhancement is $1.53 \pm 0.07^{stat} \pm 0.41^{sys} $ (without $\rho$) for minimum bias collisions.

The results are compared to different theoretical calculations \cite{Rapp, PHSD, HJXu, Yukinao}. The calculation \cite{Rapp}, which includes hadronic gas with medium modifications and radiation from the QGP, is compared to data in Fig.\ref{slope} (left) and describes the data in reasonable way. Also the PHSD \cite{PHSD} model, which includes several production mechanism as collisional broadening of vector mesons, microscopic secondary multiple-meson channels, Radiation from the QGP, is able to describe the data. The data agree with the calculations that include the in-medium modification of the vector meson in LMR.


In the IMR, compared to minimum-bias collisions, the yield from binary-scaled charm contributions from PYTHIA (with charm cross section $\sigma_{c\bar{c}}$ = 0.96mb) over-predicts the data in central collisions, which could indicate the modification of charm production in central Au + Au collisions. As shown in Fig.\ref{slope} (right), while the transverse mass slope parameters, which obtained from the transverse mass spectra\cite{NA60}, in p + p collisions are consistent with the PYTHIA charm, they are different in Au + Au, since they are higher than those in p + p collisions. This may be due to thermal radiation and/or charm modification. We also compare our results with SPS results. Our inclusive dielectron slope parameters are higher than the SPS dimuon results (charm/DY contribution subtracted) from NA60 Collaboration\cite{NA60}.

In the future, the Heavy Flavor Tracker and Muon Telescope Detector upgrades will help us to separate the charm and thermal radiation contributions.

\section{Acknowledgments}
This work was supported in part by the National Natural Science Foundation of China under contract No. 11035009, 10905085, 11275250 and the Knowledge Innovation Project of the Chinese Academy of Sciences under Grant No. KJCX2-EW-N01.


\end{document}